\documentstyle[aps,twocolumn]{revtex}
\def\eth{\hbox{$\partial$\kern-0.45em\raise0.55ex\hbox{--}}}
\def\thorn{I\kern-0.4em\raise0.35ex\hbox{\it o}}
\def\d{{\rm d}}

\newtheorem{prop}{Proposition}

\newtheorem{theo}[prop]{Theorem}

\begin{document}
\twocolumn[\hsize\textwidth\columnwidth\hsize\csname @twocolumnfalse\endcsname

\title{No Black Hole Theorem in Three-Dimensional Gravity}
\author{Daisuke Ida\dag}
\address{\dag Department of Physics, Kyoto University,
Kyoto 606-8502, Japan\\ 
{\rm Electronic address: ida@tap.scphys.kyoto-u.ac.jp}}
\date{September 8, 2000}
\maketitle
\begin{abstract}
A common property of known black hole solutions in (2+1)-dimensional gravity
is that they require a negative cosmological constant.
In this letter, it is shown that a (2+1)-dimensional gravity theory which
satisfies the dominant energy condition forbids the existence of a black hole
to explain the above situation.\\
\medskip

\medskip
\end{abstract}
]
The (2+1)-dimensional theory provides us with one of useful approaches to 
more complicated (3+1)-dimensional classical gravity
or conceptual problems in quantum gravity \cite{carlip}.
At first sight, the (2+1)-dimensional gravity looks trivial.
In particular, the vacuum Einstein equation implies that the space-time
is locally flat, corresponding to the absence of the
 gravitational radiation (Weyl tensor) in three dimensions.
However, the local distribution of matter fields has a global effect on
the outer empty space; for instance, the gravitational field
of a point particle is described by a conical space with its deficit angle 
corresponding to the mass of the particle \cite{deser84},
which causes the gravitational lens effect.
One should also 
note that the triviality of local geometry does not necessarily 
imply the triviality of the
theory itself; namely, the topological degrees of freedom plays an important role
in the theory of gravitation \cite{witten,hosoya}.
The triviality of local geometry in 
the (2+1)-gravity theory holds even if the cosmological term
is taken into account. The Einstein space is simply a space of constant curvature 
in three dimensions, so that educated relativists would not imagine that
there is a black hole solution in this theory until in 1992 Ba\~nados {\it et al.}
show that there actually exists a black hole in the locally anti-de Sitter
space \cite{BTZ,BTHZ}.
This black hole space-time, called BTZ black hole, is obtained by identifying
certain points of (the covering manifold of) the anti-de Sitter space.
A different identification makes a space-time representing
the BTZ black hole in a closed universe \cite{siino},
multiple BTZ black holes \cite{brill} or a creation of the
 BTZ black hole \cite{matschull}.
The BTZ black hole is characterized by the mass,  angular momentum and 
cosmological constant, and has almost all features of the Kerr-anti-de Sitter black hole
in the conventional four-dimensional Einstein gravity.
The BTZ black hole was shown to be also the solution of a low energy string theory
\cite{horowitz93,kaloper}.

Since the discovery of the BTZ black hole, a number of authors have attempted to find a
black hole solution in various theories in (2+1)-dimensions.
Black holes in topologically massive gravity \cite{deser82} with the negative
cosmological constant were found by Nutku \cite{nutku}.
In Einstein-Maxwell-$\Lambda$ system, a static (non-rotating) charged black hole had
been already noted in the original paper by Ba\~nados {\it et al.} \cite{BTZ}.
 Cl\'ement 
\cite{clement96} generated from the charged BTZ black hole a class of rotating charged black holes. Though rotating solutions in Einstein-Maxwell-$\Lambda$ theory seem to have infinite total mass and angular momentum,
these divergences may be cured by adding a Chern-Simons term to the action \cite{clement96}.
Black holes with a dilaton field have been discussed by many authors.
In Brans-Dicke theory, Sa {\it et al.}  found black hole solutions \cite{sa96,sa98},
and their properties were extensively studied for different Brans-Dicke parameters.
Black holes in Einstein-Maxwell-dilaton-$\Lambda$ theory were obtained by Chan and Mann 
in non-rotating \cite{chan94} and rotating \cite{chan96} cases. Other families were
given by Koikawa {\it et al.} \cite{koikawa97} and by Fernando \cite{fernando99}.
Chen \cite{chen99} also derived rotating black hole solutions in this theory
by means of the duality transformation in the equivalent non-linear $\sigma$-model.
Black holes coupled to a topological matter field \cite{carlip95}, 
conformal scalar field \cite{martinez}, Yang-Mills field \cite{brindejone},
 Born-Infeld field \cite{cataldo99} {\it etc.}
were also discussed.

Thus, many black hole solutions 
are known
Here, it might be interesting to note that all the black hole solutions listed above
require a negative cosmological constant, otherwise a certain kind of energy conditions
is violated. A typical example might be the BTZ black hole. As already mentioned,
the BTZ black hole may be constructed by making identifications in the anti-de Sitter
space. We may also consider a similar construction in the de Sitter space.
In this case, a natural procedure might be identifying two geodesic circles in
each Poincar\'e disk associated with the open chart of the de Sitter space.
The resultant space-time represents an inflating universe rather than a black hole.
The absence of black hole in this example might be due to the difference in
the causal structure of conformal infinity \cite{private}.

The purpose of this letter is to give a reason for this situation.
In particular, we will be able to answer the question: {\em ``Why the BTZ black hole
requires a negative cosmological constant?''} 
In the following, we consider the possibility of the existence of 
a black hole (in the sense of an apparent horizon)
in three-dimensional space-time
with the procedure given by Hawking \cite{hawking} in terms of the spin-coefficient
formalism \cite{spin}.

Let $(M,g)$ be a three-dimensional space-time and let $\Sigma$ be a space-like
hypersurface in $M$. Suppose that $\Sigma$ contains 
outer trapped surfaces, then there will be an
apparent horizon $H$ which is defined to be the outer boundary of the trapped region
in $\Sigma$, where the notion ``outer'' is assumed to be well-defined as in 
the case of the asymptotically flat (or anti-de Sitter) space-time.
We also assume that the apparent horizon $H$ is a smooth closed curve in $\Sigma$.
Let $m$ be a unit tangent vector of $H$, and let $n$ and $n'$ be future directed
out-going and in-going null vectors orthogonal to $H$, respectively, such that
$g(n,n')=1$.
The vectors $n$ and $n'$ are arranged such that $n-n'$ lies in $\Sigma$, which is
always possible by means of the boost transformation
$n\mapsto a^2 n$, $n'\mapsto a^{-2}n'$ by some positive function $a$.
Let us consider a local deformation of $H$ within $\Sigma$ outside the trapped region
generated by a vector field $X=e^f(n-n')$ with some smooth function $f$.
Accordingly, the null triad $\{n,n',m\}$ is extended such that the normalization
$g(n,n')=-g(m,m)=1$, $g(n,n)=g(n',n')=g(n,m)=g(n',m)=0$ 
is preserved and that
$m$ is tangent to each deformed $H$. Then, since $X$ and $Y=e^h m$ form 
holonomic base vectors on $\Sigma$ for some function $h$,
 $n$ and $n'$ are propagated such that
\begin{equation}
\delta f=\kappa-\tau+\beta=\kappa'-\tau'-\beta, \label{constraint}
\end{equation}
where Ricci rotation coefficients
\begin{eqnarray}
&&\kappa=g(m,Dn),~~\tau=g(m,D'n),~~\beta=g(n',\delta n),\nonumber\\
&&\kappa'=g(m,D'n'),~~\tau'=g(m,Dn')
\end{eqnarray}
and the differential operators
\begin{equation}
D=\nabla_n,~~D'=\nabla_{n'},~~\delta=\nabla_m
\end{equation}
are defined following the spin-coefficient formalism
in four space-time dimensions \cite{spin}.
The convergence of light rays emitted outward from each deformed $H$
 is measured by
the quantity
\begin{equation}
\rho=g(m,\delta n).
\end{equation}
In particular, $\rho=0$ holds on $H$ since $H$ will be a marginally trapped surface.
The change in $\rho$ along $X$ is derived 
by
the following equations
\begin{eqnarray}
&&D\rho-\delta\kappa=(\epsilon+\rho)\rho-(2\beta+\tau+\tau')\kappa+\phi_{++},
\label{f}\\
&&D'\rho-\delta\tau=-\epsilon'\rho-\kappa\kappa'-\tau^2+\rho\rho'-\phi_{+-}-\Pi
\label{cf},
\end{eqnarray}
where
\begin{eqnarray}
&&\epsilon=g(n',Dn),~~\epsilon'=g(n,D'n'),
~~\rho'=g(m,\delta n'),\nonumber\\
&&\phi_{++}=\phi(n,n),~~\phi_{+-}=\phi(n,n'),~~\Pi=R/6
\end{eqnarray}
with the trace-free part of the Ricci tensor
$\phi=-{\rm Ric}+(R/3)g$.
Subtracting Eq. (\ref{cf}) from Eq. (\ref{f}), we obtain the equation
\begin{eqnarray}
e^{-f}{\mathcal L}_X\rho&=&\delta(\kappa-\tau)-(2\beta+\tau+\tau')\kappa
+\kappa\kappa'+\tau^2\nonumber\\
&&{}+\phi_{++}+\phi_{+-}+\Pi\nonumber\\
&=&\delta(\delta f-\beta)+(\kappa-\tau)^2+\phi_{++}+\phi_{+-}+\Pi \label{key}
\end{eqnarray}
on $H$, where Eq.~(\ref{constraint}) has been used.
Now suppose that there is a positive cosmological constant $\Lambda>0$
and that the stress-energy tensor $T$
satisfies {\em the 
 dominant energy condition:
(i) $T(W,W)\ge 0$, and (ii) $T(W)$ is non-space-like,
for every time-like vector $W$}. 
Then, the Einstein equation ${\rm Ric}-(R/2)g+\Lambda g=-8\pi T$ 
leads to 
the inequalities
\begin{equation}
\phi_{++}\ge 0,~~\phi_{+-}+\Pi>0.
\end{equation}
The term $\delta(\delta f-\beta)$ in the last line of the Eq. (\ref{key}) can be made
zero by 
appropriately choosing the function $f$;
in fact, parametrizing
$H$ by the proper length $s\in [0,{\rm Length}(H))$, 
such a function $f$ can be explicitely written as
\begin{equation}
f=\int^s\beta\d s-\left(\frac{\oint\beta\d s}{\oint\d s}\right)s.
\end{equation}
Then, the last line of the Eq.~(\ref{key}) is positive definite, 
${\mathcal L}_X\rho> 0$.
This implies that there is an outer trapped surface outside $H$,
which contradicts the assumption that $H$ is the outer boundary of such surfaces.
Hence, we obtain the following  no black hole theorem:
\begin{theo}
Let $(M,g)$ be a three-dimensional space-time subject to the Einstein equation
${\rm Ric}-(R/2)g+\Lambda g=-8\pi T$ with $\Lambda>0$. If the stress-energy tensor
$T$ satisfies the dominant energy condition, then $(M,g)$ contains no apparent horizons.
\label{nobh}
\end{theo}

This explains why black hole solutions require a negative cosmological constant.
Strictly speaking, we can only say that there is no non-degenerate
apparent horizon ($\rho=0$, ${\mathcal L}_X\rho\ne 0$) in the case of $\Lambda=0$,
however, the presence of matter fields such as the dilaton or Maxwell field will exclude
even degenerate horizons. 

Thus, a black hole in (2+1)-gravity requires
negative energy such as a negative cosmological constant.
This implies the breakdown of the predictability in certain 
three-dimensional theories.
As in four space-time dimensions, we may consider the Oppenheimer-Snyder model of
the gravitational collapse. The homogeneous disk of dust will collapse 
to a central point and a naked conical singularity will be left.
This picture of gravitational collapses will remain unchanged 
unless the negative cosmological constant is added.
Even in the case of the non-symmetric
gravitational collapse of  gauge fields or  scalar fields,
there will not form a black hole, so that when a singularity is formed, such a
singularity will be naked.

We have disscussed the existence problem of apparent horizons,
while the black hole is often defined by the event horizon.
Since the theorem \ref{nobh} relies on the local analysis, we cannot argue the global
structure of space-time such as an event horizon. 
An important exception is the stationary case; we can replace ``apparent horizons''
with ``stationary event horizons'' in the theorem \ref{nobh}, 
since it is known that these coincide in this case.

The author would like to acknowledge helpful discussions with
Prof.\ H. Sato, Prof.\ K. Nakao and Dr.\ M. Siino.
He would also like to thank Dr. S. Higaki for careful reading of the manuscript.
He was supported by JSPS Research Fellowships for Young Scientists,
and this research was supported in part by the Grant-in-Aid for Scientific
Research Fund (No. 4318).

\end{document}